\documentstyle[aps,preprint]{revtex}

\newcommand{\be}{\begin{equation}}
\newcommand{\ee}{\end{equation}}
\newcommand{\ba}{\begin{eqnarray}}
\newcommand{\ea}{\end{eqnarray}}

\newcommand{\grgl}{\:\hbox to -0.2pt{\lower2.5pt\hbox{$\sim$}\hss}
           {\raise3pt\hbox{$>$}}\:}
\newcommand{\klgl}{\:\hbox to -0.2pt{\lower2.5pt\hbox{$\sim$}\hss}
           {\raise3pt\hbox{$<$}}\:}

\draft

\begin{document}
\preprint{HD--THEP--97--29,~~~
 \\ SOGANG--HEP 223/97}

\title{Quantization of spontaneously broken gauge theory
        \\ based on the BFT--BFV Formalism}

\author{Yong-Wan Kim\footnote{\noindent e-mail:~ywkim@physics.sogang.ac.kr}
and 
Young-Jai Park\footnote{e-mail:~yjpark@ccs.sogang.ac.kr}}

\address{Department of Physics and Basic Science Research Institute\\
Sogang University, C.P.O.Box 1142, Seoul 100-611, Korea\\
and\\
Institut  f\"ur Theoretische Physik\\
Universit\"at Heidelberg\\
Philosophenweg 16, D-69120 Heidelberg, Germany}

\maketitle

\begin{abstract}
\noindent We quantize the spontaneously broken abelian U(1) Higgs model
by using the improved BFT and BFV formalisms. 
We have constructed the BFT physical fields,
and obtain the first class observables including the Hamiltonian
in terms of these fields.
We have also explicitly shown that there are exact form invariances 
between the second class and first class quantities.
Then, according to the BFV formalism, we have derived 
the corresponding Lagrangian having U(1) gauge symmetry.
We also discuss at the classical level how one easily gets the first
class Lagrangian from the symmetry-broken second class Lagrangian.
\end{abstract}

\pacs{PACS:~11.10.Ef, 11.15.Ex, 11.15.-q}


\section{Introduction}

Many of the fundamental theories of modern physics can be considered
as descriptions of dynamical systems subjected to constraints.
The foundations for Hamiltonian quantization of these constrained systems
have been established by Dirac \cite{Di}. By requiring the 
strong implementation of second class constraints, however,
this method implies Dirac brackets, 
whose non-canonical structure may pose serious problems
on operator level. This makes it desirable to embed the second class
theory into a first class one in which the commutator relations remain 
canonical. 

An example is provided by the Higgs model with spontaneous
symmetry breakdown \cite{AL} whose quantization
is usually carried out in the so called ``unitary'' gauge. As is well known,
in this gauge the model is 
a purely second class system characterized by
two sets of the second class constraints \cite{GR,BRR1}.
The required strong implementation of these constraints
leads to non-polynomial field dependent Dirac brackets. 
As mentioned above, 
one can circumvent the problems associated with this
non-polynomial dependence by turning this system into a first class one with 
a usual Poisson bracket structure in an extended phase space
and implementing the first class constraints on the physical states. 

A systematic procedure for achieving this has been given by Batalin and
Fradkin (BF) \cite{BF} in the canonical formalism, and applied to various
models obtaining the Wess-Zumino (WZ) action \cite{FIK}. 
In particular, this analysis explicitly carried out for the above 
Higgs model \cite{BRR1}. 
However, it is already proved that the construction 
of the first class Hamiltonian in the BF framework
is  non-trivial even in the abelian case because of the field dependence
on the constraint algebra.
In this case the only
weakly involutive first class Hamiltonian is obtained
after the fifth iteration, and thus it does not appear          
particularly to be suited for treating non-abelian cases. 

A more systematic and transparent approach for this iterative procedure, 
called Batalin--Fradkin--Tyutin (BFT) formalism 
when combined with the BF one, 
has been developed by Batalin and Tyutin \cite{BT}. 
This procedure has been applied to
several interesting models \cite{BR,AD}, 
where the iterative process is terminated after two steps. 
In general, it has been, however, still difficult to apply this BFT
formalism to the nonabelian case \cite{BBG}. 
On the other hand, we have recently improved the BFT
formalism by introducing the novel concept of the BFT physical fields 
constructed in the extended phase space \cite{KK}
in order to construct the strongly involutive observables
including the Hamiltonian. 
This modified version of the BFT method has been 
successively applied for only finding the 
first class Hamiltonian of several nontrivial 
nonabelian models \cite{BBN,PP,KR}. 
In particular, the origin of the second class constraints
of the Higgs model comes from the spontaneous symmetry breaking effects,
while that of non-abelian Proca model \cite{BBN,PP} due to the existence
of the explicitly symmetry broken mass term. Therefore, it is very
interesting to analyze the Higgs model having the different origin,
which is phenomenologically important.

In this letter we shall revisit the spontaneously broken abelian U(1) 
Higgs model by following the constructive procedure
based on the improved BFT version \cite{KK}.
In section 2, we convert the second class 
constraints into a first class ones, and  construct in section 3 
the BFT physical fields in the extended phase space
corresponding to the original fields in the usual phase space,
following the improved BFT formalism.
We then systematically obtain all observables containing 
the first class Hamiltonian as functionals of the BFT physical fields
showing the form invariances between the second class and first 
class quantities.
In section 4, through the standard path integral quantization 
established by Batalin, Fradkin and Vilkovsky (BFV) \cite{FV,HEN}, 
we derive the gauge invariant Lagrangian. 
In section 5, we suggest a novel path at the classical level how one 
can easily get this Lagrangian from the original second class one by
simply replacing the original fields with the BFT ones through
an additional relation without following usually complicated path
integral quantization. This new method will be possible to analyze
the realistic non-abelian Higgs models.
We summarize in section 6.

\section{BFT construction of first class constraints}

Consider the abelian U(1) Higgs model in the unitary gauge \cite{GR,BRR1},
\be\label{2.1}
{\cal L}_u = -\frac{1}{4} F_{\mu\nu}F^{\mu\nu} 
           + \frac{1}{2} g^2 (\rho + v)^2 A_\mu A^\mu 
           + \frac{1}{2} \partial_\mu \rho \partial^\mu \rho           
           + V(\rho),\ee
where the subscript ``$u$'' stands for the unitary gauge, the Higgs 
potential is $V(\rho)={1 \over 2} \mu^2 (\rho+v)^2-{\lambda \over 4}
(\rho+v)^4$, and the field strength tensor $F_{\mu\nu}=\partial_\mu
A_\nu - \partial_\nu A_\mu$. The momenta canonically conjugate to
$A^0, A^i$ and $\rho$ are given by $\pi_0=0, \pi_i=F_{i0}$,
and $\pi_\rho=\dot \rho$, respectively. 
We have thus one primary constraint
\be\label{2.2}
\Omega_1=\pi_0\approx 0.\ee
The canonical Hamiltonian density associated with the Lagrangian 
(\ref{2.1}) is given by
\be\label{2.3}
{\cal H}_c = {1 \over 2} \pi^2_i + {1 \over 2} \pi^2_\rho 
             + {1 \over 4} F_{ij} F^{ij}
             - {1 \over 2} g^2 (\rho+v)^2 \left( (A^0)^2 - (A^i)^2 \right)
             - A^0 \partial^i \pi_i
             + {1 \over 2} (\partial_i \rho)^2             
             - V(\rho).\ee
Persistency in time of $\Omega_1$ leads to one further (secondary) 
constraint
\be\label{2.4}
\Omega_2 = \partial_i \pi^i + g^2 (\rho+v)^2 A^0 \approx 0.\ee
Then, the constraints $\Omega_i$ in Eqs. (\ref{2.2}) and (\ref{2.4}) 
consist of a second class system because we have the nonvanishing
Poisson brackets
\be\label{2.5}
\Delta_{ij}(x,y) \equiv 
    \{\Omega_i(x), \Omega_j(y)\}=-g^2(\rho+v)^2\epsilon_{ij}\delta^3(x-y),\ee 
where $\epsilon_{12}=\epsilon^{12}=1$.

We now convert this second class system defined by the commutation relations
(\ref{2.5}) to a first class one at the expense of introducing additional
degrees of freedom. According to the BFT method \cite{BT}, 
we first introduce 
two auxiliary fields $\Phi^i$ corresponding to $\Omega_i$  
with the Poisson brackets
\be\label{2.6}
\{\Phi^i(x),\Phi^j(y)\}=\omega^{ij}(x,y),\ee
where we are free to make a choice
\be\label{2.7}
\omega^{ij}(x,y)=\epsilon^{ij}\delta^3(x-y).\ee
The first class constraints $\tilde \Omega_i$ 
are then constructed as a power series in the auxiliary fields:
\be\label{2.8}
\tilde \Omega_i = \sum^{\infty}_{n=0}\Omega^{(n)}_i ;                  
             ~~\Omega^{(0)}_i=\Omega_i,\ee
where $\Omega^{(n)}_i$ are homogeneous polynomials in the auxiliary fields 
${\Phi^j}$ of degree $n$, to be determined by the requirement that the
first class constraints $\tilde \Omega_i$ be strongly involutive:
\be\label{2.9}
\left\{\tilde\Omega_i(x),\tilde\Omega_j(y)\right\}=0.\ee
Since $\Omega^{(1)}_i$ are linear in the auxiliary fields, 
we could make the ansatz
\be\label{2.10}
\Omega^{(1)}_i=\int d^3y X_{ij}(x,y)\Phi^{j}(y).\ee
Then, substituting (\ref{2.10}) into (\ref{2.9}) leads to the following
relation 
\be\label{2.11}
\int d^3 z d^3 z'X_{ik}(x,z)\omega^{kl}(z,z')X_{jl}(z',y)
               =-\Delta_{ij}(x,y).\ee
For the choice of (\ref{2.7}), Eq. (\ref{2.11}) has a solution
\be\label{2.12}
X_{ij}(x,y)=\left( \begin{array}{cc}
             g^2(\rho+v)^2 & 0 \\
               0 &  1 
               \end{array} \right) \delta^3(x-y).\ee
Substituting (\ref{2.12}) into (\ref{2.10}) as well as (\ref{2.8}),
and iterating this procedure one finds the strongly involutive first class
constraints to be given by 
\ba\label{2.13}
\tilde \Omega_1 &=& \Omega_1 + g^2(\rho+v)^2\Phi^1, \nonumber \\
\tilde \Omega_2 &=& \Omega_2 + \Phi^2. \ea
This completes the construction of the first class constraints
in the extended phase space.

\section{Construction of first class observables}

The construction of the first class Hamiltonian $\tilde {\cal H}$ can be
done along similar lines as in the previous case of the constraints. 
However, we shall
follow here a somewhat different path \cite{KK} by using the novel property
that any functional
${\cal K}(\tilde{\cal F})$ of the first class fields $\tilde{\cal F}=
(\tilde A^\mu, \tilde \pi_\mu, \tilde \rho, \tilde \pi_\rho)$ 
corresponding to the original fields ${\cal F}=
(A^\mu, \pi_\mu, \rho, \pi_\rho)$ will also be first class. i. e., 
\be\label{3.1}
\tilde {\cal K}({\cal F};\Phi)={\cal K}(\tilde {\cal F}).\ee
This leads us to the identification 
$\tilde {\cal H}_c={\cal H}_c(\tilde {\cal F})$. To do this, we should 
first construct first class ``physical fields" $\tilde {\cal F}$ in the 
extended phase space, which are obtained as a power series in the auxiliary
fields $\Phi^j$ by requiring them to be strongly involutive:
$\{\tilde \Omega_i,\tilde {\cal F}\}=0$. 
Expressions of the strongly involutive $\tilde {\cal F}$ are given by
\ba
\tilde A^\mu &=&(A^0+{1 \over g^2(\rho+v)^2}\pi_\theta, 
                 A^i+\partial^i\theta) \label{3.2}, \\
\tilde \pi_\mu &=&(\pi_0+g^2(\rho+v)^2\theta,\pi_i)\label{3.3}, \\
\tilde \rho &=&\rho,\label{3.4} \\
\tilde \pi_\rho &=& \pi_\rho + 2g^2(\rho+v)A^0\theta.\label{3.5}\ea
Here, for the later convenience, 
we have identified the auxiliary fields $\Phi^i$
as a canonically conjugate pair $(\theta,\pi_\theta)$ by choosing
\ba
\Phi^i=(\theta,\pi_\theta),\nonumber \ea
which satisfy the symplectic structure (2.6) with the choice of 
Eq. (2.7).
In order to understand the meaning of these BFT fields,
let us now consider the Poisson brackets between the BFT fields
in the expended phase space.
From the relations of Eqs. (\ref{3.2}--\ref{3.5}),
one can easily calculate the Poisson brackets of the abelian 
Higgs model as follows
\ba\label{3.11}
&& \{ \tilde A^0(x), \tilde A^i(y) \} 
    = {1 \over g^2(\tilde\rho+v)^2} \partial^i_x \delta^3 (x-y),
   \nonumber \\
&& \{\tilde A^0(x), \tilde \pi_\rho (y) \} 
    = -{2 \tilde A^0 \over (\tilde\rho+v)}  \delta^3(x-y),  
   \nonumber \\
&& \{\tilde A^i(x), \tilde \pi_j (y) \} 
    = \delta^i_j \delta^3(x-y),
   \nonumber \\
&&  \{\tilde \rho (x), \tilde \pi_\rho (y)\}
    = \delta^3(x-y).   
\ea
Note that these Poisson brackets in the extended phase space
have the form invariance as compared with the Dirac brackets in the
original phase space, and moreover,
if we take the limit of $\Phi^i=(\theta,\pi_\theta) 
\rightarrow 0$, the brackets (\ref{3.11}) are nothing but the
usual Dirac brackets of the abelian Higgs model \cite{GR}.
On the other hand, we observe 
that the first class constraints (\ref{2.13}) can
be written in terms of the BFT physical fields $\tilde {\cal F}$ as
\ba\label{3.6}
\tilde \Omega_1 &=& \tilde \pi_0, \nonumber \\
\tilde \Omega_2 &=& \partial^i \tilde \pi_i + g^2(\tilde\rho+v)^2 
                    \tilde A^0,\ea
and these constraints also have the form invariance with
the second class constraints $\Omega_i$ in Eqs. 
(\ref{2.2}) and (\ref{2.4}).

Correspondingly, we take the 
first class Hamiltonian density $\tilde {\cal H}_c$ to be given by 
the second class one (\ref{2.3}), expressed in terms of the physical fields:
\be\label{3.7}
\tilde {\cal H}_c = {1 \over 2} \tilde \pi^2_i 
             + {1 \over 2} \tilde \pi^2_\rho 
             + {1 \over 4} \tilde F_{ij} \tilde F^{ij}
             - {1 \over 2} g^2 (\tilde \rho+v)^2 
                   \left( (\tilde A^0)^2 - (\tilde A^i)^2 \right)
             - \tilde A^0 \partial^i \tilde\pi_i
             + {1 \over 2} (\partial_i \tilde \rho)^2 
             - V(\tilde \rho),\ee
and, by construction, $\tilde H_c = \int d^3x ~\tilde{\cal H}_c$ 
is automatically strongly involutive
\be\label{A.10}
\{\tilde \Omega_i, \tilde H_c \} = 0.
\ee

It seems appropriate to comment on our strongly involutive 
Hamiltonian (\ref{3.7}) derived by using the improved BFT formalism.
Making use of (\ref{3.2}--\ref{3.5}) and (\ref{3.7}), we may rewrite 
$\tilde {\cal H}_c$ in the form
\ba\label{3.33}
\tilde {\cal H}_c = {\cal H}_c + \Delta{\cal H}
      - {\pi_\theta\over g^2(\rho+v)^2}\tilde \Omega_2,  \ea
where $\Delta {\cal H}$ is given by 
\ba
\Delta{\cal H} &=& 2 g^2(\rho+v)\theta A_0
                  \left(\pi_\rho+g^2(\rho+v)\theta A_0\right)
                  -g^2(\rho+v)^2\partial^i\theta(A_i+{1\over2}\partial_i
                   \theta) \nonumber \\
&+& {\pi^2_\theta \over 2g^2(\rho+v)^2}. \nonumber
\ea
If we construct our strongly involutive Hamiltonian along similar BFT lines 
as in the case of the constraints, we obtain 
this Hamiltonian (\ref{3.33}) after only
second iterations (see the Appendix A), 
while the weakly involutive 
Hamiltonian in ref. \cite{BRR1} 
derived by using the BF formalism
is obtained after the fifth iteration 
in spite of the abelian case and has rather complicated additional terms 
given in Eq. (2.29) of ref. \cite{BRR1}. 
Any Hamiltonian weakly equivalent to (\ref{3.7}), however, 
describes the same physics 
since the observables of the first class formulation must be 
first class themselves, and thus these two Hamiltonians are
equivalent to each other.
Therefore, we can add to $\tilde {\cal H}_c$
any terms freely proportional to the first class constraints. 
In particular, if we choose the simplest Hamiltonian density
among infinite equivalent ones:
\be\label{3.8}
\tilde {\cal H}'_c =  {\cal H}_c 
                     + \Delta {\cal H},\ee
then this naturally generates the Gauss law constraint 
$\tilde \Omega_2$
\be\label{3.9}
\{\tilde \Omega_1, \tilde {\cal H}'\} = \tilde \Omega_2,~~~
\{\tilde \Omega_2, \tilde {\cal H}'\} = 0,\ee
and it will be proved to be useful through the following discussion 
as well as the next section.

If we consider this Hamiltonian (\ref{3.8}), 
the form-invariant Hamilton's equations
of motion for the physical BFT fields are found to be read
\ba\label{3.12}
\dot{\tilde A^0} &=& \partial^i \tilde A^i 
                   + {2 \over (\tilde\rho+v)}\tilde A^i\partial^i\tilde\rho
                   - {2 \over (\tilde\rho+v)}\tilde\pi_\rho \tilde A^0,
               \nonumber \\
\dot{\tilde\pi}_0 &=& \tilde\Omega_2, \nonumber \\
\dot{\tilde A^i} &=& \tilde\pi_i + \partial^i \tilde A^0, \nonumber \\
\dot{\tilde \pi}_i &=& - \partial^j \tilde F_{ij} 
                       - g^2 (\tilde\rho + v)^2 \tilde A^i,
               \nonumber \\
\dot{\tilde \rho} &=& \tilde\pi_\rho, \nonumber \\
\dot{\tilde\pi}_\rho &=& -g^2(\tilde\rho+v)\left( 
                    (\tilde A^0)^2 + (\tilde A^i)^2 \right)
                      + \partial^2_i \tilde\rho + V(\tilde\rho)',\ea
where $V(\tilde\rho)'={\partial \over \partial\tilde\rho}V(\tilde\rho)$. 
But, if one try to derive the equations of motion from 
the strongly involutive  Hamiltonian (\ref{3.7}), 
one can only obtain the weak relations since 
as an example the Hamilton's equation of motion for 
$\tilde A^i$ is reduced to be  
\ba
\dot{\tilde A^i} = \tilde\pi_i + \partial^i \tilde A^0 -
\partial^i_x \left( {1 \over g^2(\tilde\rho+v)^2} 
                    \tilde {\Omega}_2 \right). \nonumber 
\ea

As a result, these relations (\ref{3.12}) together with 
Eqs. (\ref{3.6}) and (\ref{3.9})
confirm the form invariances between the second class quantities 
in the original phase space and the corresponding first class
ones in the extended phase space.

\section{Corresponding first class Lagrangian}

In order to interpret the results presented at the previous sections
from the Lagrangian point of view,
let us apply the BFV quantization scheme \cite{FV,HEN}
to the first class system described by Eqs. (\ref{3.6}) and (\ref{3.8}).
We first introduce ghosts 
${\cal C}^i$, antighosts ${\cal P}^i$ and new auxiliary fields
$q^i$ with their canonically conjugate momenta $\overline{\cal P}^i,
\,\overline{\cal C}_i$ and $p_i$ such that 
\be\label{4.1}
[ {\cal C}^i, \overline{\cal P}_j ]
      = [ {\cal P}^i, \overline{\cal C}_j ]=[q^i,p_j]=i\delta^i_j
\delta^3(x-y),\ee
where the subscript $i,j=1,2$, due to having the two constraints in Eq. 
(\ref{3.6}), and from now on we will use the commutators instead of 
the Poisson brackets. 
The nilpotent BRST charge $Q$ and the ferminonic gauge
fixing function $\Psi$ are then given by
\ba\label{4.2}
Q &=& \int d^3x \left( {\cal C}^i \tilde\Omega_i 
                       + {\cal P}^i p_i \right), \nonumber\\
\Psi &=& \int d^3x \left( \overline{\cal C}_i \chi^i
                       + \overline{\cal P}_i q^i \right),\ea
where $\chi^i$ are gauge fixing functions satisfying the condition
$det\{\chi^i, \tilde\Omega_j\}\neq 0$ \cite{HEN,FS}. 
The total unitarizing Hamiltonian is then given by
\be\label{4.3}
H_T=H_m+{1 \over i}[\Psi,Q].\ee
Since we have the involutive relations (\ref{3.9}), the minimal Hamiltonian
$H_m$ is nothing but
\be\label{4.4}
H_m=\tilde H'_c + \int d^3x \overline{\cal P}_2{\cal C}^1.\ee

The corresponding quantum theory is now defined by the extended 
phase space functional integral
\ba\label{4.5}
&& {\cal Z}_{\rm I} = \int {\cal D}\mu_{\rm I} e^{iS_{\rm I}}; \nonumber \\
&& S_{\rm I}=\int d^3x \left( 
            \pi_\mu\dot{A}^\mu + \pi_\rho\dot{\rho}
            + \pi_\theta\dot{\theta} 
            +  p_i \dot{q}^i 
             +  {\cal C}_i \dot{\overline {\cal P}\,}^i 
             +  {\cal P}_i \dot{\overline {\cal C}\,}^i
             - {\cal H}_T \right); \nonumber \\
&& {\cal D}\mu_{\rm I} = {\cal D}A^\mu{\cal D}\pi_\mu
                 {\cal D}\rho{\cal D}\pi_\rho
                 {\cal D}\theta{\cal D}\pi_\theta 
                 {\cal D}q^i{\cal D}p_i
                 {\cal D}{\cal C}^i {\cal D}\overline{\cal P}_i 
                 {\cal D}{\cal P}^i {\cal D}\overline{\cal C}_i.
\ea
According to the BFV formalism \cite{FV,HEN}, 
${\cal Z}_{\rm I}$ is independent of the
choice of the gauge fixing functions $\chi^i$. By choosing the proper 
$\chi^i$, which do not include the 
ghosts, antighosts, and auxiliary fields and their conjugate momenta, and 
taking the limit of $\beta \rightarrow 0$ after rescaling the field 
variables as $\chi^i \rightarrow \chi^i/\beta,~ p_i \rightarrow
\beta p_i$, and $\overline{\cal C}_i \rightarrow \beta \overline{\cal C}_i$,
one can integrate out all the ghost, antighosts and auxiliary variables 
in the partition function. As a result, one obtains
\be\label{4.6}
 {\cal Z}_{\rm II} = \int {\cal D}\mu_{\rm II} 
                     \prod_{i,j} \delta(\tilde\Omega_i)\delta(\chi^j)    
                     det[\chi,\tilde\Omega]e^{iS_{\rm II}},
\ee
where $S_{\rm II}$ is the action in the extended phase space
\be\label{4.7}
S_{\rm II}=\int d^4x \left(
            \pi_\mu\dot{A}^\mu + \pi_\rho\dot{\rho}
            + \pi_\theta\dot{\theta} - {\cal H}'_c\right),\ee
and ${\cal D}\mu_{\rm II}$ is the measure containing all the fields 
and their conjugate momenta except for the ghosts, antighosts 
and auxiliary fields in
the measure ${\cal D}\mu_{\rm I}$. Note that this form of the partition
function $S_{\rm II}$ coinsides with the Faddeev-Popov formula \cite{FP}.

Now, let us perform the momentum integrations to obtain the configuration
partition function. In general, if we choose the Faddeev-Popov type
gauges in which the gauge fixing conditions $\chi^i$ only depend on the 
configuration space variables, one can 
easily carry out the momentum integrations
in the partition function (\ref{4.6}).

The $\pi_0$ integration is trivially performed by exploiting the delta
function $\delta(\tilde\Omega_1)$, and after exponentiating the remaining 
delta function $\delta(\tilde\Omega_2)$ in terms of a Fourier variable
$\xi$ as $\delta(\tilde\Omega_2)
=\int{\cal D}\xi \exp (-i\int d^4x \xi\tilde\Omega_2)$ and 
transforming $A^0 \rightarrow A^0+\xi$, the integration over the momenta
$\pi_\theta, \pi_\rho$ and $\pi_i$ leads to
\be\label{4.8}
 {\cal Z} = \int {\cal D}A^\mu {\cal D}\rho
                 {\cal D}\theta{\cal D}\xi
                 \prod_i\delta(\chi^i)det[\chi,\tilde\Omega]det
                 (g(\rho+v))
                 e^{iS_{\rm GI}},
\ee
where
\be\label{4.9}
S_{\rm GI} = \int d^4x \left( -{1 \over 4}F_{\mu\nu}F^{\mu\nu}
             + {1 \over 2}[
                \partial_\mu+ig(A_\mu+\partial_\mu\theta)](\rho+v)
          [ \partial^\mu-ig(A^\mu+\partial^\mu\theta)](\rho+v)
             + V(\rho)\right), \ee
and the $\xi$ field in the measure is an artifact which would be 
removed if we take the gauge fixing function $\chi^i$ explicitly.
This action $S_{\rm GI}$ is now gauge invariant under the transformation
\ba\label{4.91}
A_\mu \rightarrow A_\mu + \partial_\mu \Lambda, ~~
\theta \rightarrow \theta - \Lambda,~~
\rho \rightarrow \rho.\ea
This completes our analysis on the BFV quantization scheme.

It only  remains to establish the equivalence 
between the above gauge invariant action 
and the well-known U(1) Higgs model. 
By defining the complex scalar field $\phi(x)$ as
\be\label{4.10}
\phi(x)=\frac{1}{\sqrt{2}} \left(\rho (x)+v \right) e^{-ig\theta (x)}\ee
with the BFT field $\theta$ playing the role of the Goldstone boson,
and replacing the Jacobian factor
${\cal D}\theta {\cal D}\rho \, det\left(g(\rho+v)\right)$ 
in the measure part with ${\cal D}\phi{\cal D}\phi^*$,
we can easily rewrite the partition function (\ref{4.8}) with 
the action (\ref{4.9}) as follows 
\ba\label{4.11}
 {\cal Z}_{\rm F}&=&\int {\cal D}A^\mu {\cal D}\phi^* {\cal D}\phi
         \prod_i\delta(\chi^i) det\{\chi,\tilde\Omega\} 
         e^{iS_{\rm F}};\nonumber\\
 S_{\rm F}&=& \int d^4x \left( -{1 \over 4}F_{\mu\nu}F^{\mu\nu} 
                  + (D_\mu\phi)^* (D^\mu\phi)
                  + \mu^2 \phi^* \phi 
                  - \lambda (\phi^* \phi)^2 \right),\ea
where $D_\mu=\partial_\mu-igA_\mu$ is the covariant derivative.
As a result, we have arrived 
at the well-known U(1) Higgs model, which describes
the interaction of the abelian gauge fields $A^\mu$ with the complex scalar
fields $\phi$, through the BFT--BFV construction.

\section{BFT Lagrangian formulation}

Let us now consider in this section the similar economical method
to simply obtain the first class Lagrangian (\ref{4.9}) at the
classical level. 
It consists in gauging the Lagrangian (\ref{2.1}) by making 
the substitution $A^\mu \rightarrow \tilde A^\mu$ and $\rho \rightarrow
\tilde\rho$.  
The spatial components $\tilde A^i$ 
among the vector potential components
contain only the fields of the configuration space
as in Eq. (\ref{3.2}),
and already take the usual form of the
gauge transformation,  i.e.,
$\tilde A^i \rightarrow A^i + \partial^i\theta$.
However, since the time component
$\tilde A^0$ contains the term of the momentum
field $\pi_\theta$ as in Eq. (\ref{3.2}), 
we should first replace this term
with some ordinary field before carrying out the above substitution.
In order to incorporate the $\tilde A^0$ field at this stage, we use 
an additional relation, which has not been recognized up to now. 

From the useful property (\ref{3.1}) and the definition of $\pi_i$, 
we observe the following relation:
\ba\label{5.1}
\tilde \pi_i &=& \partial_i\tilde A_0-\partial_0\tilde A_i \nonumber\\
        &=& \partial_i\left(A_0+{1 \over g^2(\rho+v)^2}\pi_\theta\right)
            -\partial_0(A_i+\partial_i\theta).\ea
On the other hand, another form of $\tilde \pi_i$ 
is already given in Eq. (\ref{3.3}) as follows
\be\label{5.2}
\tilde \pi_i = \pi_i = \partial_i A_0 - \partial_0 A_i.\ee
Comparing this with Eq. (\ref{5.1}), we see that the following
additional relation
should be maintained for the consistency all the times
\be\label{5.3}
\partial^0\theta={1 \over g^2(\rho+v)^2}\pi_\theta,\ee
which make it possible to directly replace the second term of 
$\tilde A^0$ with $\partial^0\theta$. 
By making use of this relation,
we can now rewrite the $\tilde A^0$ 
as the usual form of the gauge transformation as follows
\be\label{5.4}
\tilde A^0 = A^0 + \partial^0 \theta. \ee
Note that the Hamilton's equations (\ref{3.12}) can be also 
used to confirm the relation (\ref{5.3}).
Therefore, gauging the original Lagrangian (\ref{2.1}), i.e., making the
substitution
\ba\label{5.5}
\tilde A^\mu \rightarrow A^\mu + \partial^\mu \theta,~~
\tilde \rho \rightarrow \rho,  \ea
we have directly arrived at the same first class Lagrangian 
(\ref{4.9}) at the classical level,
\be\label{5.6}
{\cal L}(\tilde A^\mu,\tilde \rho)=\tilde {\cal L}(A^\mu,\theta,\rho)
={\cal L}_{\rm GI},\ee
which is already obtained through the standard path integral procedure
in the previous section.

\section{Summary}

In this letter, we have quantized the spontaneously broken abelian U(1)
Higgs model, which is a phenomenologically interesting and simple toy model,
through the BFT--BFV quantization procedure.

First, according to the improved version \cite{KK} of the BFT formalism, 
we have constructed the BFT physical fields,
and proved that the Poisson brackets between these BFT fields
naturally contain the structure of the Dirac brackets \cite{GR}
in the original phase space for the abelian U(1) Higgs model 
as like in Eq. (\ref{3.11}),
while maintaining the form invariance in the extended phase space.

Second, we have shown that the strongly involutive 
first class Hamiltonian (\ref{3.7}) is directly obtained by replacing the 
second class fields with the first class BFT ones. 

Third, after directly obtaining the above Hamiltonian and 
by choosing the simplest involutive Hamiltonian ${\cal H}'_c$ among 
equivalent infinite ones,
we have shown that this $\tilde {\cal H}'_c$ 
in Eq. (\ref{3.8}) naturally generates the Gauss law constraint, 
and also gives the strong Hamilton's equations of motion (\ref{3.12}).
Moreover, we have also shown that there are the exact form invariances
between the second class and first class quantities in Eqs. (\ref{3.6},
\ref{3.7}) and (\ref{3.12}), which give us the deep physical
meaning of the BFT fields 
when we embed a second class system into first class by using the
BFT construction. 

Fourth, we have carried out the BFV quantization procedure
in order to interpret
the results of the Hamiltonian embedding of the abelian U(1) Higgs model
from the Lagrangian point of view, and constructed the gauge invariant
Lagrangian corresponding to the first class Hamiltonian.

Fifth, by using the additional relation (\ref{5.3}), 
we have newly shown that one can directly obtain 
the first class Lagrangian from the second class one 
by just replacing the original fields
with the BFT ones at the classical level, similar to the case of 
the Hamiltonian.
In particular, this kind of the BFT Lagrangian construction will be 
powerful for the analysis of
the non-abelian cases, where the non-abelian extension of the 
$\tilde A^0$ remains to be weakly involutive to the usual 
gauge transformation. 

In conclusion, we have shown that the improved version \cite{KK}
of the BFT formalism
is more economical than the previous BFT versions \cite{BR,AD,BBG} 
including the BF one \cite{BRR1} 
by explicitly analyzing the abelian U(1) Higgs model. 
We hope that this powerful BFT formalism with the additional 
relation, which we first used here, 
will improve the transparency of the analysis in the nonabelian cases
which are realistic and phenomenological models 
related to the spontaneously broken symmetry.

\section*{Acknowledgment}
We would like to thank M.--I. Park and K. D. Rothe 
for helpful discussions, and 
the Institut f\"ur Theoretische Physik for their
warm hospitality. The present study was partly supported by  
the Korea Research Foundation for (1996) overseas fellowship (Y.--W. Kim), 
the KOSEF--DFG Exchange Program (Y.--J. Park), 
and the Basic Science Research Institute Program, Ministry of Education,
1997, Project No. BSRI--97--2414.

\end{document}